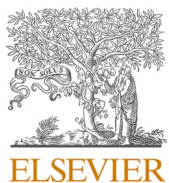
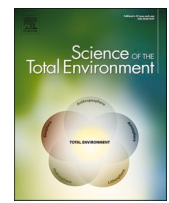

# A ten-year historical analysis of urban $PM_{10}$ and exceedance filters along the Northern Wasatch Front, UT, USA

Callum E. Flowerday [a], Rebekah S. Stanley [a], John R. Lawson [b,c], Gregory L. Snow [d], Kaitlyn Brewster [a], Steven R. Goates [a], Walter F. Paxton [a], Jaron C. Hansen [a,*]

[a] *Department of Chemistry and Biochemistry, Brigham Young University (BYU), Provo, UT, USA, 84602*
[b] *Bingham Research Center, Utah State University, Vernal, UT 84078, USA*
[c] *Dept. of Mathematics and Statistics, Utah State University, Logan, UT 84322, USA*
[d] *Department of Statistics, Brigham Young University (BYU), Provo, UT 84602, USA*

## HIGHLIGHTS

- Analysis shows no correlation between Great Salt Lake size and urban PM10 levels.
- Winds from south/west playas and deserts are key sources of urban dust.
- Urban dust metallic composition indicates minimal health risk from toxic metals.
- Windrose plots highlight dust origins tied to playa and desert regions near the lake.
- Hazard quotients and cancer risks for dust metals are below health concern levels.

## GRAPHICAL ABSTRACT

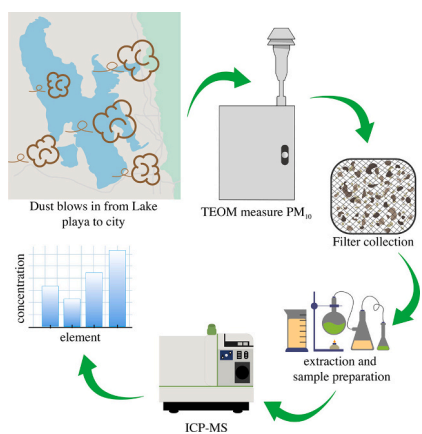

ABSTRACT

The Great Salt Lake (Utah, USA) is reducing in size, which raises several ecological concerns, including the effect of an increasing area of dry playa exposed by the retreating lake. This study focuses solely on concerns about the toxicity of metals in the dust blowing off the playa. Although considerable efforts have been made to understand aeolian dust in urban areas along the Wasatch Front, located just east and south of the Great Salt Lake, there is still a need to consolidate existing research and to conduct a compositional analysis of the dust found in these urban areas. We investigated the dust reaching urban monitoring sites around the Great Salt Lake that are managed by the Utah Division of Air Quality. By analyzing historical data, we found that the decrease in the Great Salt Lake's surface area has not led to a statistically significant increase in dust events in urban areas. Windrose plots align with prior research, indicating that heightened dust levels in urban areas coincide with winds originating from the south or west, passing over identified playas and deserts such as the Milford Flats, Sevier Dry Lake, Tule Dry Lake, Great Salt Lake Desert, Dugway Proving Grounds, and the West Desert of Utah.

* Corresponding author.
  *E-mail address:* jhansen@chem.byu.edu (J.C. Hansen).

Metallic compositional analysis of urban dust was used to evaluate potential health risks associated with the dust using the hazard quotient, air regional screening levels, and cancer risk methods. This analysis revealed no significant increase in concentrations of toxic metals. However, this is not to preclude a risk of dust-related health concerns, especially due to pre-existing arsenic and lead levels.

*Synopsis:* This study provides insights into dust-related health risks and environmental impacts in Utah, analyzing dust exposure from shrinking Great Salt Lake.

## 1. Background

The Great Salt Lake (GSL) has been shrinking for the past 35 years, reducing in size from 1046.0 miles$^2$ (2709 km$^2$) on January 1, 2015, to 940.5 miles$^2$ (2436 km$^2$) on January 1, 2022. This reduction has exposed 105.5 miles$^2$ of playa that was previously submerged. As this playa dries out, it becomes a potential source of dust that may be carried by the wind into urban areas, primarily to the south and east of the GSL, leading to reduced visibility and potential health concerns.

Research indicates that the deposition of aeolian (windblown) dust on mountainous snowpacks accelerates their melting, which adversely affects Utah's essential water supply throughout the year (Lang et al., 2023; Carling et al., 2020; Skiles et al., 2018; Carling et al., 2012; Nicoll et al., 2020; Hall et al., 2021). This windblown dust is characterized as particulate matter with a diameter of 10 μm or smaller (PM$_{10}$), routinely measured in accordance with the National Ambient Air Quality Standards (NAAQS) established by the United States Environmental Protection Agency (EPA) (Lawrence and Neff, 2009; United States Environmental Protection Agency, 2022a).

These particles have the potential for carrying heavy metals that possibly originate from the playa surrounding the GSL (Lawrence and Neff, 2009), which raises concerns about health risks associated with dust inhalation (Han et al., 2017; Jeong and Ra, 2021). Various analytical methods, including gas chromatography–mass spectrometry (GC–MS) (Han et al., 2017; Jeon et al., 2001), X-ray fluorescence (Han et al., 2017; Hahnenberger and Perry, 2015; Thorsen et al., 2017), and inductively coupled plasma mass spectrometry (ICP-MS) (Carling et al., 2020; Carling et al., 2012; Jeong and Ra, 2021; United States Environmental Protection Agency, 2016; United States Environmental Protection Agency, 2005; United States Environmental Protection Agency, 2022b; Putman et al., 2022; Goodman et al., 2019), have been employed to quantify the concentrations of heavy metals in PM$_{10}$.

Considerable research has been conducted to trace the origins of dust along the Wasatch Front in Utah, USA (Lang et al., 2023; Carling et al., 2020; Nicoll et al., 2020; Hahnenberger and Perry, 2015; Putman et al., 2022; Goodman et al., 2019; Bouton et al., 2020; Hahnenberger and Nicoll, 2012; Hahnenberger and Nicoll, 2014; Steenburgh et al., 2012; Munroe et al., 2023). A significant portion of this dust has been identified as originating from playas or the Great Salt Lake Desert (GSLD). Lang et al. (Lang et al., 2023), utilizing backwind trajectories, determined that 23 % of the dust deposited in the snow on the Wasatch Mountains originated from the GSL playa, whereas 45 % came from the GSLD or the playas of Sevier Dry Lake and Tule Dry Lake in southern Utah. Carling et al. (Carling et al., 2020) used strontium isotope ratios in deposited dust to reveal that the GSL playa contributes 5 % of dust along the southern Wasatch Front and 30–34 % of dust along the northern Wasatch Front.

Putman et al. (Putman et al., 2022) concluded from strontium isotope ratios on the dust fraction of <63 μm that most of the dust from these playas, especially the coarsest dust, was deposited outside urban areas. They found that much of the dust measured in these areas originated from local soil material or activities, such as industrial processes, mining, oil refining, and agriculture, with a smaller contribution from regional playas. They also noted the presence of metals such as As, La, V, Pb, Tl, and Ni but concluded that these metals are linked to pollution generated by local industries and soil contamination. Hahnenberger et al. (Hahnenberger and Perry, 2015) studied soil and dust samples from Sevier Dry Lake and discovered compositional differences between soil and dust samples for minor soil elements, although major soil elements (Si, Al, Fe, Ca, Na, K, Mg, and Ti) showed similarities. They also found that part of the dust in the Salt Lake metropolitan area could be traced to the Sevier Dry Lake.

In a separate study, Hahnenberger et al. (Hahnenberger and Nicoll, 2012) determined that dust storms affecting the Salt Lake City metropolitan area primarily occur in the spring, during late afternoon when westerly winds prevail. They identified Tule Dry Lake, Sevier Dry Lake, GSLD, and Milford Flats as dust sources, noting that 60 % of dust originated from playas classified as barren land cover. 75 % of dust from vegetated land cover originated from the area in the Milford Flats that were burned in the 363,052-acre, 2007 Milford Flat Fire (Hahnenberger and Nicoll, 2014). Goodman et al. (Goodman et al., 2019), through mass balance calculations, estimated that up to 90 % of dust along the Wasatch Front originates from playas, with source locations matching those listed above. Steenburgh et al. (Steenburgh et al., 2012) observed a general decline in dust days from 1930 to 2012, with emission sources identified as Sevier Dry Lake, Milford Valley (which includes the Milford Flats), West Desert of Utah, Escalante Desert, and the Great Basin and Mojave Deserts of Nevada. Nicoll et al. (Nicoll et al., 2020) identified Milford Valley, Sevier Dry Lake, Tule Dry Lake, GSLD, and the Dugway Proving Grounds as major dust sources. Munroe et al. (Munroe et al., 2023) identified similar regions as sources of dust using a network of dust samplers, dust deposition rates, and backwind trajectories.

Other topics, such as environmental justice (Grineski et al., 2024) and sources of cyanobacteria and cyanotoxins (Metcalf et al., 2023) in dust, have also been studied. Metcalf et al. (Metcalf et al., 2023) found no clear relation between the presence of cyanotoxins in airborne samples and those found in adjacent lakebed samples, suggesting that such airborne toxins may originate from diffuse sources rather than a single point source. A recent study evaluated the oxidative potential of dust generated from the GSL playa (Attah et al., 2024), measuring 47 metals in samples from the area. Arsenic (As) levels were found to exceed the EPA's soil residential regional screening levels (RSLs) for all playas across Utah, and As and lithium (Li) for GSL dust were above the RSLs. Additionally, lead (Pb), copper (Cu), manganese (Mn), iron (Fe), and aluminum (Al) were found to be associated with the oxidative potential of dust from the GSL.

Acknowledging that there are other concerns associated with the shrinking of the GSL, this study leverages historical data to specifically assess the correlation between the surface area of the GSL and the quantity of dust observed in urban areas. Additionally, we identify dust sources through wind rose plots, conduct a compositional analysis on dust collected in urban areas around the GSL, and evaluate potential health risks associated with the dust using the hazard quotient, air regional screening levels, and cancer risk methods.

## 2. Method

### 2.1. Surface area versus dust analysis

Particulate matter can be classified as PM$_{10}$ and PM$_{2.5}$. PM$_{10}$ is defined as particulate matter that is 10 microns (μm) or less in diameter (Utah Department of Air Quality, 2024a). PM$_{2.5}$ is defined as fine, inhalable particles or droplets with a diameter of 2.5 μm or smaller (Utah Department of Air Quality, 2024b). PM$_{10}$ may consist of soot, metals, salt, and dust; major sources may include vehicles, wood-burning, wildfires or open burns, industry, dust from construction sites, landfills, gravel pits, agriculture, and open lands. The EPA established different National Ambient Air Quality Standards (NAAQS) for $PM_{10}$ and $PM_{2.5}$, in part, because the larger $PM_{10}$ particles are largely filtered out in the nose and throat, whereas $PM_{2.5}$ particles can penetrate deeply into the lungs, resulting in more serious health effects. The EPA NAAQS specify the maximum amount of PM permissible in outdoor air. The standard for $PM_{10}$ is currently a 24-hour average of 150 μg/m$^3$ for a 24-hour period. The standard for PM2.5 is 35 μg/m$^3$ for the 98th percentile averaged over 3 years. A further classification is coarse fraction of PM. Coarse particulate matter (PM) refers to particles ranging from 2.5 to 10 μm in size, typically defined as the difference between $PM_{10}$ and $PM_{2.5}$. Even though this fraction is generally recognized, the EPA has not established an exposure standard for it.

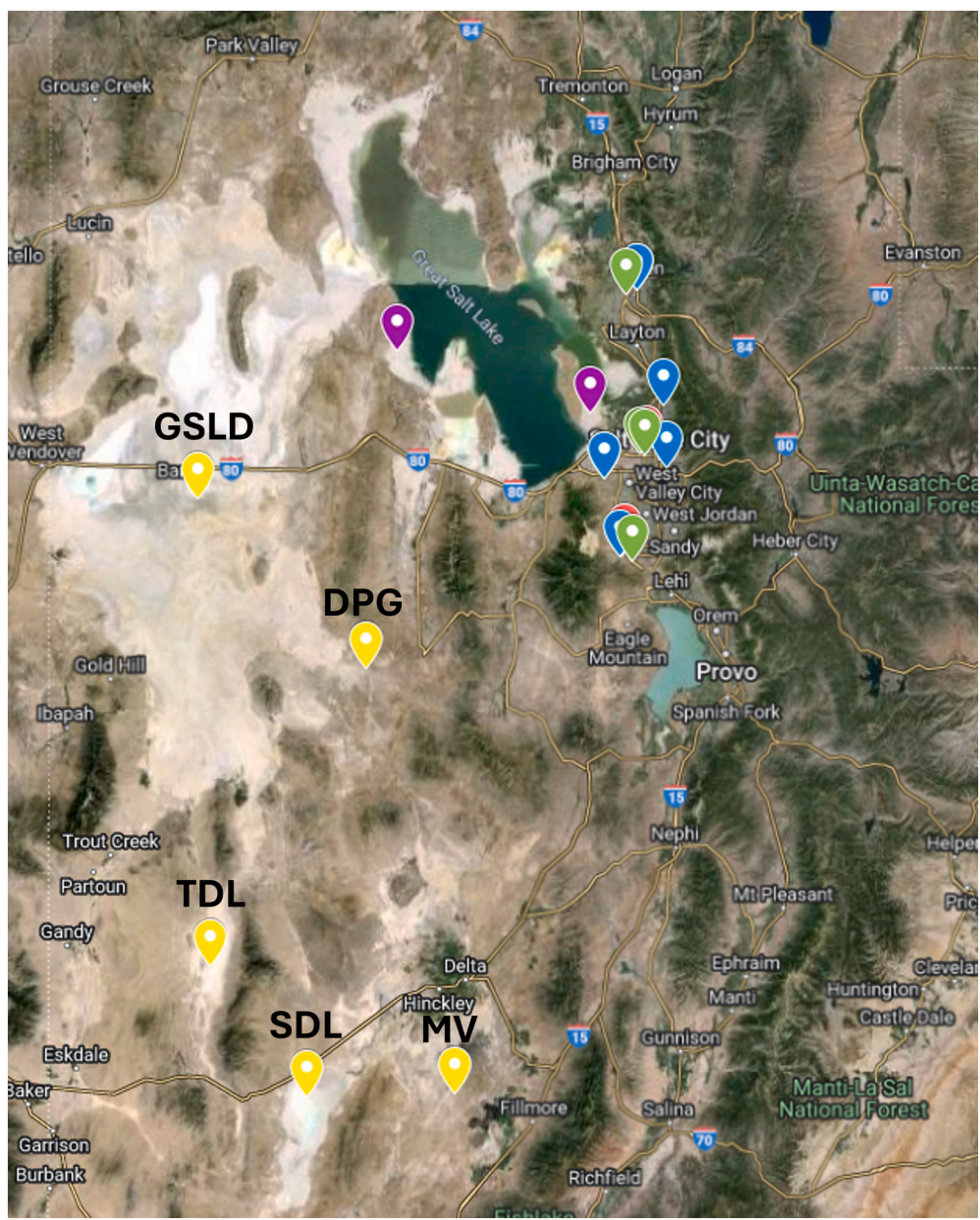

**Fig. 1.** Map of Utah, USA. Blue pins – Locations of $PM_{10}$ monitoring sites from which $PM_{10}$ concentrations were collected to detect correlations between dust and the surface area of the GSL (Maps, 2024). Red pins – Locations of $PM_{10}$ monitoring sites from which filters were collected and then analyzed. Green pins – Meteorological stations used to get meteorological data (Synoptic Weather, 2024). Purple pins – partial region of the exposed GSL playa. Yellow pins – Identified regions of $PM_{10}$ contributions to dust measured in urban regions including the Great Salt Lake Desert (GSLD), Dugway Proving Grounds (DPG), Tule Dry Lake (TDL), Sevier Dry Lake (SDL), and Milford Valley (MV) (Maps, 2024).

Lake-depth and surface-area data used in this study were sourced from the United States Geological Survey (USGS) Utah Water Science Center (United States Geological Survey, 2021; Tarboton, 2023; Tarboton, 2024; Tarboton, 2017; United States Geological Survey, 2005; United States Geological Survey, 2006). The EPA provided quality-assured $PM_{2.5}$ and $PM_{10}$, both averaged hourly and daily for five urban sampling sites (United States Environmental Protection Agency, 2024a; United States Environmental Protection Agency, 2023a). These data were critical when tracking the $PM_{10}$ concentration trends in relation to changes in the surface area of the GSL. Furthermore, they were employed to assess the correlation between the GSL's surface area and $PM_{10}$ concentrations.

Our analysis focused on five representative urban sites: Bountiful (EPA AIRS Code: 490110004; address: 171 West 1370 North, Bountiful, UT), Hawthorne Elementary (EPA AIRS Code: 490353006, address:

**Table 1**

PM$_{10}$ filters acquired from the Utah Division of Air Quality with corresponding concentrations, GSL surface area, and predominant wind direction for that day.

| Filter # | Site | Date | PM$_{10}$ concentration (μg/m$^3$) | Notes | Lake SA (miles$^2$) | PM$_{2.5}$ concentration (μg/m$^3$) | % PM$_{2.5}$ |
|---|---|---|---|---|---|---|---|
| 0634111 | Herriman (H3) | 04/23/2022 | 8.1 | Baseline | 940.4831 | 1.9 | 23 |
| 0634365 | Herriman (H3) | 05/31/2022 | 6.2 | Baseline | 940.4831 | 1.5 | 24 |
| 0634535 | Herriman (H3) | 06/17/2022 | 118.8 | 3rd from max | 926.744 | 19.8 | 17 |
| 0634539 | Herriman (H3) | 06/20/2022 | 15.2 | Baseline | 926.744 | 4.1 | 27 |
| 0634559 | Utah Tech Center (EQ) | 06/17/2022 | 146.7 | 2nd from max | 926.744 | 24.1 | 16 |
| 0634560 | Utah Tech Center (EQ) | 06/18/2022 | 145.7 | 3rd from max | 926.744 | 29.8 | 20 |
| 0634563 | Utah Tech Center (EQ) | 06/20/2022 | 9.3 | Baseline | 926.744 | 3.0 | 32 |
| 9610827 | Utah Tech Center (EQ) | 09/07/2020 | 162.7 | Exceedance | 1001.206 | 38.4 | 24 |
| 9610829 | Utah Tech Center (EQ) | 09/09/2020 | 10.6 | Baseline | 1001.206 | 2.7 | 25 |
| 7672706 | Herriman (H3) | 12/20/2017 | 229.4 | Exceedance | 1064.408 | 22.9 | 10 |
| 7672719 | Magna (MG) | 12/20/2017 | 170.0 | Exceedance | 1064.408 | 21.7 | 13 |
| 7672720 | Magna (MG) | 12/21/2017 | 9.0 | Baseline | 1064.408 | 2.1 | 23 |
| 2917265 | Herriman (He) | 04/14/2015 | 256 | Exceedance | 1116.213 | N/A | – |
| 2917266 | Herriman (He) | 04/15/2015 | 6 | Baseline | 1116.213 | N/A | – |
| 2917234 | Ogden (O2) | 04/14/2015 | 332 | Exceedance | 1116.213 | 32.9 | 10 |
| 2917235 | Ogden (O2) | 04/15/2015 | 5 | Baseline | 1116.213 | 2.5 | 50 |

EPA AIRS code and addresses: H3 – 490353013 (14058 Mirabella Drive, Herriman), EQ – 490353015 (240 North 1950 West, Salt Lake City), MG – 490351001 (2935 S 8560 W, Magna), He – 490353008 (12885 5600 W, Herriman), O2 – 490570002 (228 32nd St, Ogden).

1675 South 600 East, Salt Lake City, UT), Herriman (EPA AIRS Code: 490353013 address: 14058 Mirabella Drive, Herriman) Magna (EPA AIRS Code: 490351001 address: 2935 South 8560 West, Magna) and Weber (EPA AIRS Code: 490570002, address: 228 East 32nd Street, Ogden City). The locations of these three sites are shown in Fig. 1.

PM$_{10}$, PM$_{2.5}$, and their difference (PM coarse fraction) were plotted against the surface area of the GSL and against the date. Curves were fitted to the plots with regression splines to highlight any general trends. The square root of each class of PM was also plotted, which partially de-skews the concentrations and dampens the effect of the few extreme datapoints. Regression models were fit to the PM measures and surface area, both straight line and estimated curves from splines, to determine if there were any relationships. Although these could be statistically significant, because of the large dataset, statistical significance does not mean that the effect is large enough to be of practical significance (Ambaum, 2010). Furthermore, a fit of data can meet significance criteria but not be meaningful, as may be the case here, because we do not find a correlation with known sources or mechanisms.

We also conducted historical dust event analysis using hourly-averaged PM$_{10}$ data for Hawthorne (490353006, years collected: 2022–2023), Ogden (490570002, years collected: 2012–2015), and Herriman (490353013, years collected: 2023). Data qualifiers, provided with the data, were used to filter out firework events (qualifier code: IH), US wildfire events (qualifier code: IT), Canadian wildfire smoke (qualifier code: IF), outliers (qualifier code: 5), and field issues (qualifier code: 3). Points that had a qualifier code for high winds (IJ) were left in the dataset.

*2.2. Meteorological analysis: visibility, wind speed, and wind direction*

Archived meteorological observations were obtained from Synoptic Weather (Synoptic Weather, 2024; Horel et al., 2002). Where available, values for visibility and wind direction and speed were obtained. If any of these data were not available for a given filter site, data were obtained from the nearest weather station. Surface-level meteorological analyses were obtained from the Weather Prediction Center (National Oceanic and Atmospheric Administration, 2024). The US National Weather Service (USNWS) (US National Weather Service, 2024) defines a dust event as having a visibility of <0.25 miles (402 m) for 3 h. Events were identified from raw time series by creating a Boolean time series by thresholding visibility. From this, another Boolean series was generated to identify a flipped bit, after which the cumulative sum of the flipped-bit series was computed to assign a new event cluster number to each consecutive group in the Boolean series. After removing all non-dust event timestamps, cluster numbers were used to determine the start and end of dust events, thereby allowing computation of the length of dust events. All events less than 3 h long were removed to match the NWS definition (US National Weather Service, 2024) of dust events being longer than 3 h long. A final sorting of dust events determined whether they were independent events or a single event as any dust event must be at least 3 h from a distinct event or it was not deemed independent. Dust events were confirmed with an hourly-averaged PM$_{10}$ concentration cutoff of 150 μg/m$^3$.

*2.3. Dust compositional analysis*

PM$_{10}$ filters, supplied by the Utah Division of Air Quality (UDAQ), underwent compositional dust analysis; these filters spanned the years 2015–2022 from five distinct urban areas bordering the Great Salt Lake Region. Fig. 1 illustrates the locations of the sampling sites. Sixteen filters, exhibiting varying PM$_{10}$ concentrations detailed in Table 1, were chosen for analysis: 8 exceedance filters and 8 baseline filters. Selection criteria included filters exceeding standards or approaching the National Ambient Air Quality Standards (NAAQS) exceedance limit of 150 μg/m$^3$ in the past ten years; however, there were no exceedance filters in 2014. UDAQ's policy of withholding exceedance filters for analysis that are less than three years old led to the selection of three other filters that approach the exceedance standard. The set comprises of eight high-concentration filters, defined as filters with >140 μg/m$^3$ of PM$_{10}$ and eight low-concentration filters, defined as filters with <16 μg/m$^3$ of PM$_{10}$, to serve as background references. The latter were collected within a few days of the high-concentration filters to maintain similar collection conditions. Analysis encompassed twelve elements of interest: As, Ba, Cd, Co, Cr, Cu, Hg, Mn, Ni, Pb, V, and Zn—many listed as Hazardous Air Pollutants by the EPA (United States Environmental Protection Agency, n.d.).

All sample extraction, preparation, and analysis procedures adhered to EPA standard operating (United States Environmental Protection Agency, 2016; United States Environmental Protection Agency, 2005; United States Environmental Protection Agency, 2022b) procedures for filter analysis. The dust sample was extracted from the filter via sonication preparation, applicable to the type of filters provided by the state, according to EPA 68-D-00-264. These were then diluted to a 2 % nitric acid solution for analysis on an inductively coupled plasma-mass spectrometer (ICP-MS, PerkinElmer, Nexion 300× with an annual Perkin Elmer service plan). The instrument passed all validation methods defined in Perkin Elmer daily check protocols before analysis began and met all passing criteria during the analysis (See Figure 24-1 (United States Environmental Protection Agency, 2005)). Seven method blanks and seven method spike samples were prepared over three days by the same extraction method as above (EPA 821-R-16-006). In addition, interference checks were prepared and run (according to the NATTS TAD (United States Environmental Protection Agency, 2022b)). Three internal standards (Y, Sc, Tb) were used to monitor samples and standards for matrix effects and instrument stability.

All dust filters and interference checks (United States Environmental Protection Agency, 2022b) were analyzed in a single day, whereas method blanks and spikes (United States Environmental Protection Agency, 2016) were analyzed on three separate days, as directed by the EPA guidelines. Each run followed the same calibration and continuing quality control check procedures. In addition, instrument tuning procedures were conducted daily as directed by manufacturer guidelines. Daily calibrations were verified by high standard analysis and initial calibration verification with a quality control standard. Calibration blanks and quality control standards were also rerun every ten samples and at the end of the runs for each day. Passing criteria can be found in EPA-68-D-00-264. Method blank and spike samples were used to calculate method detection limits according to EPA 821-R-16-006 by

$$MDL_s = t_{(n-1, 1-\alpha=0.99)} S_s \qquad (1)$$

where $MDL_s$ is the method detection limit based on spiked samples, $t_{(n-1, 1-\alpha=0.99)}$ is Student's t-value appropriate for the single-tailed 99th percentile, and $S_s$ is the sample standard deviation of the replicates of spike samples and

$$MDL_b = \overline{X} + t_{(n-1, 1-\alpha=0.99)} S_b \qquad (2)$$

where $MDL_b$ is the method detection limit based on blank samples, $\overline{X}$ is the mean of method blank results, $t_{(n-1, 1-\alpha=0.99)}$ is Student's t-value appropriate for the single-tailed 99th percentile, and $S_b$ is the sample standard deviation of the replicates of blank samples. These MDLs can be found in supplemental material. After instrument analysis, metal concentration in the air was calculated from the dust filter sample data according to

$$C = \left[ \left( :g\frac{metal}{L} \right) \left( Digestion\ Volume \left( \frac{L}{filter} \right) \right) \right] \Big/ V_{std} \qquad (3)$$

where $C$ is the concentration of the metal, :g metal/L is the metal concentration defined in section 15.1 of the procedure, the digestion volume is 0.050 L as defined by the procedure, and $V_{std}$ is the standard air volume pulled through the filter in m$^3$, which was provided for each filter by UDAQ. Uncertainties, at the 95 % confidence interval, associated with the quantification of metals in the dust samples were combined in quadrature to produce an uncertainty associated with each element. These uncertainties, seen in Table 7, originated from measurement in the ICP-MS (<1 % uncertainty for each element, per EPA passing criteria), measurement of the standard volume of air passed through the filter (5 % uncertainty), and from sample preparation and dilution (2 % uncertainty). All waste produced was disposed of according to EPA guidelines.

### 2.4. Health risk assessment

#### 2.4.1. Air regional screening levels

Air Regional Screening Levels (RSLs) (United States Environmental Protection Agency, 2023b) are used to assess contaminant risk at superfund sites. During the initial screening process, the maximum concentration of each chemical measured at a superfund site is compared to the respective RSL. Those chemicals with values below the RSL are removed from the risk assessment, whereas chemicals with values equal to or above the RSL become Chemicals of Potential Concern (COPC). Only COPC are taken through the rest of the risk assessment process, including health risk assessments and possible remediation measures. RSL values for the elements analyzed were obtained from the EPA (United States Environmental Protection Agency, 2024b) website;

**Table 2**
Metal toxicity levels as defined by the Environmental Protection Agency (EPA) (United States Environmental Protection Agency, 2023b; United States Environmental Protection Agency, 2024b; Agency for Toxic Substances and Disease Registry (ASTSDR), 2021), National Ambient Air Quality Standard (NAAQS) (United States Environmental Protection Agency, 2022a), Occupational Safety and Health Administration (OHSA) (Administration, n.d.), who report an 8-hour exposure standard, and the California Office of Environmental Health Hazard Assessment (OEHHA) (California Office of Environmental Health Hazard Assessment, 2024), who set an acute, an 8-hour, and chronic exposure standard.

| Element | EPA RfC standard (µg/m$^3$) | NAAQS | OSHA standard (µg/m$^3$) | OEHHA standard (µg/m$^3$) | | |
|---|---|---|---|---|---|---|
| | | | | Acute | 8-hour | Chronic |
| As | 0.015 | – | 10 | 0.200 | 0.015 | 0.015 |
| Ba | 0.5[c] | – | 500 | – | – | – |
| Cd | 0.01[c] | – | 5 | – | – | 0.02 |
| Co | 0.02[a] | – | – | – | – | – |
| Cr | – | – | 1000[b] | – | – | – |
| Cu | – | – | – | 100[b] | – | – |
| Hg | 0.3 | – | – | 0.6 | 0.06 | 0.03 |
| Mn | 0.05 | – | 5000 | – | 0.17 | 0.09 |
| Ni | 0.014[c] | – | – | 0.2 | 0.06 | 0.014 |
| Pb | – | 0.15[b] | – | 15 | – | – |
| V | 0.1[c] | – | 500 | 30 | – | – |
| Zn | – | – | 5000[b] | – | – | – |

[a] RfC currently under development, chronic exposure value obtained from Agency for Toxic Substances and Disease Registry (ATSDR) toxicological profile.
[b] Value used in HQ calculations in lieu of EPA RfC.
[c] Values taken from EPA regional screening level resident ambient air table.

those with a Hazard Quotient of 1 or greater and a Cancer Risk of at least $1.0 \times 10^{-6}$ are listed in Table 7.

#### 2.4.2. Hazard quotient

One method for determining whether the concentrations of metals in dust pose a noncancerous health risk involves calculating the hazard quotient (HQ) (Jeong and Ra, 2021; Agency for Toxic Substances and Disease Registry (ASTSDR), 2022). An HQ >1 indicates potential for adverse health effects, whereas an HQ <1 suggests minimal to no risk of such effects. The HQ is computed from Eq. (4), where the EF-adjusted Air Concentration (µg/m$^3$) (AAC) is divided by the reference concentration (RfC) for each specific metal, also in units of µg/m$^3$. The AAC is calculated from Eq. (5), where Contaminant Concentration (µg/m$^3$) (C) is multiplied by the Exposure Factor (unitless) (EF). The EF is calculated with Eq. (6), where Exposure Frequency (F) is multiplied by the Exposure Duration (ED) and divided by the Averaging Time (AT). The reference concentrations for the tested metals are outlined in Table 2 and were obtained from the EPA Integrated Risk Information System (IRIS) (United States Environmental Protection Agency, 2024c) site and the RSL tables mentioned above (United States Environmental Protection Agency, 2024b). In cases where an EPA RfC value was unavailable, data for metal toxicity levels were taken from other regulatory agencies as noted in Table 2.

$$HQ = \frac{AAC}{RfC} = \frac{adjusted\ air\ concentration}{reference\ concentration} \qquad (4)$$

$$AAC = (C \times EF) \qquad (5)$$

$$EF = \frac{(F \times ED)}{AT} \qquad (6)$$

For these noncancer hazard quotient calculations, the risk was averaged over 33 years, which is the 95th percentile of resident occupancy time (ASTSDR, 2020). The Exposure Frequency and Exposure Duration were determined by the average number and duration of dust events at the monitored sites.

Table 3
EPA Inhalation Unit Risk (IUR) values from the Integrated Risk Information System (IRIS) (United States Environmental Protection Agency, 2024c) and Air Regional Screening Level (RSL) (United States Environmental Protection Agency, 2024b) tables.

| Element | IUR $(\mu g/m^3)^{-1}$ |
|---|---|
| As | $4.3 \times 10^{-3}$ |
| Ba | Lack of studies |
| Cd | $1.8 \times 10^{-3}$ |
| Co | $9.0 \times 10^{-3}$ |
| Cr | Not Available |
| Cu | Not Available |
| Hg | Inadequate Data |
| Mn | No Human Data |
| Ni | $2.4 \times 10^{-4}$ |
| Pb | Inadequate Human Data |
| V | Not Available |
| Zn | Inadequate Data |

### 2.4.3. Cancer risk

The Cancer Risk (CR) (Agency for Toxic Substances and Disease Registry (ASTSDR), 2022) associated with a contaminant is calculated separately from the noncancer hazard quotient and involves Inhalation Unit Risks (IRUs) for each individual element (Agency for Toxic Substances and Disease Registry (ASTSDR), 2022). IURs were obtained from the EPA's Integrated Risk Information System (IRIS) (United States Environmental Protection Agency, 2024c) site and the RSL tables and are listed in Table 3 (United States Environmental Protection Agency, 2024b).

A CR was calculated for each element with an IUR by Eq. (7), where AAC $(\mu g/m^3)$ is multiplied by the IUR in $(\mu g/m^3)^{-1}$ and then multiplied by the ratio of the Age-Specific Exposure Duration (ED) in years divided by the Lifetime in Years (LY), which for cancer risk is 78 years (Agency for Toxic Substances and Disease Registry (ASTSDR), 2022).

$$CR = (AAC \times IUR) \times \left(\frac{ED}{LY}\right) \quad (7)$$

If the CR is $>1.0 \times 10^{-6}$, the element poses a cancer risk and further toxicology analysis should be performed. For these calculations the 95th percentile of resident occupancy time of 33 years (ASTSDR, 2020) was used for the ED, meaning these calculations are for an individual residing in the monitored area for 33 years.

## 3. Results and discussion

### 3.1. Surface area versus dust analysis

Fig. 2 shows the annually averaged PM, shown as $PM_{2.5}$, $PM_{10}$, and coarse fraction, compared to the surface area of the GSL. There has been a slight decline or stagnation in PM concentrations in all three PM fractions since the year 2000, despite a decline in the surface area of the lake. $PM_{2.5}$ 98 % and $PM_{10}$ 2nd max plots, found in the supplemental material, also show a slight general decline or a stagnation, despite a decrease in GSL surface area.

Fig. 3A is an example of plots that depict the available $PM_{10}$ concentrations plotted over time. Fig. 3B is an example of the square root of $PM_{10}$ concentrations plotted over time. This was done to partially de-skew the measures, dampening the effect of a few extreme points. Fig. 3C is an example of a correlation plot illustrating the relation between the surface area of the GSL and $PM_{10}$ concentrations. Corresponding plots were made for all five representative urban sites (Bountiful, Hawthorne, Herriman, Magna, and Weber) for $PM_{10}$, $PM_{2.5}$, and the coarse fraction ($PM_{diff}$); these can be found in the supplemental materials.

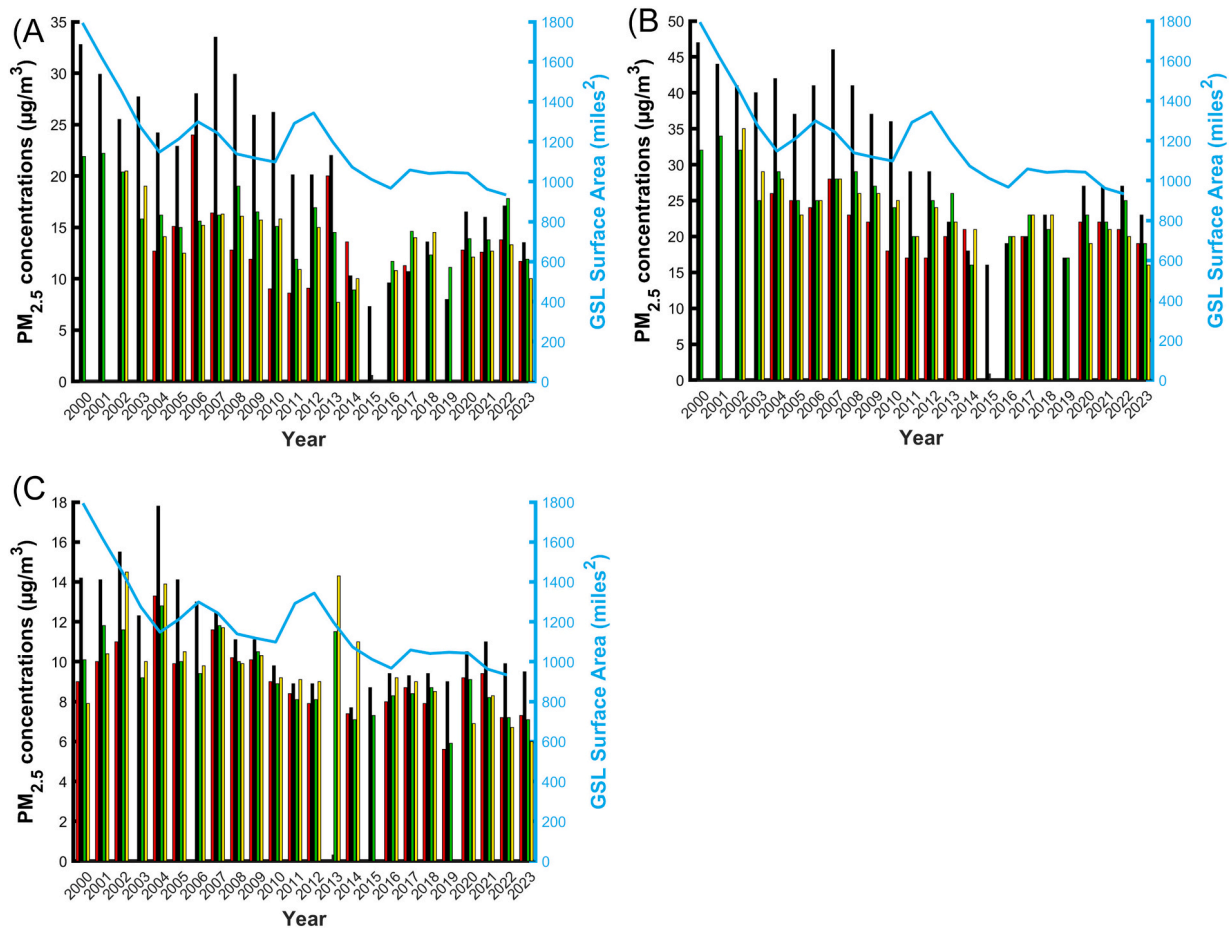

**Fig. 2.** Annually averaged PM plotted compared to annually averaged GSL surface area (blue). PM is divided into (A) coarse fraction (B) $PM_{10}$, and (C) $PM_{2.5}$. All PM data are plotted where available. For each year, data for four counties are plotted in a repeating order: Davis County (red), Salt Lake County (black), Utah County (green), Weber County (yellow).

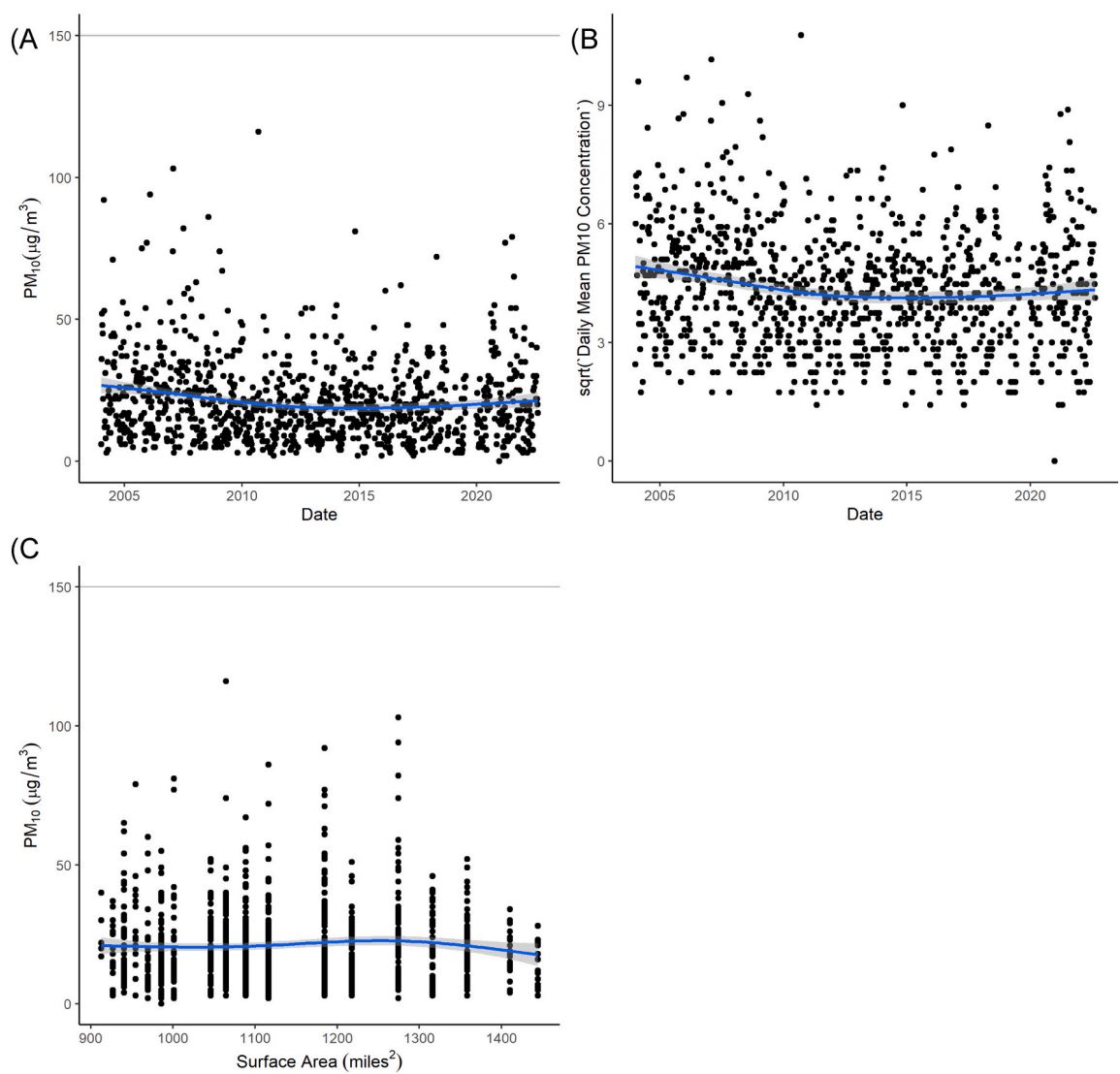

**Fig. 3.** (A) Example of $PM_{10}$ plotted overtime for Bountiful. The grey line at 150 μg/m³ shows exceedance events. The blue line represents fitted regression model, with 95 % confidence interval shown as grey shading. (B) Example of square root of $PM_{10}$ plotted over time for Bountiful. The blue line represents fitted regression model, with 95 % confidence interval shown as grey shading. (C) Example of a correlation plot illustrating the relation between the GSL's surface area and $PM_{10}$ concentrations for Bountiful. The blue line represents fitted regression model, with 95 % confidence interval shown as grey shading.

The surface of the GSL has consistently declined from 2015 to 2022 exposing 105.5 mile² (273 km²) of previously covered playa. By contrast, the average concentration of $PM_{10}$ has maintained a steady level during the same period. The correlation plots underscore that these two trends exhibit minimal correlation, if any, showing little support to the idea that reduction in the surface area of the GSL increases $PM_{10}$ dust levels in the populated urban areas surrounding the lake. This lack of correlation was consistent across all five representative urban sites that surround the eastern and southern regions of the GSL.

This analysis was performed for the five representative urban sites: Bountiful, Hawthorne, Herriman, Magna, and Weber. A relation that produces a p-value <0.05 is generally considered statistically significant. The p-values for the spline fits test a null hypothesis that the curved fit (such as those seen in the plots in Fig. 3) does not fit any better than a straight line. For each of the five representative urban sites, the spline curve shows a significant non-zero relation with $p < 0.001$; a0 p0-value of <0.05 means that the flexible curve fits better than a straight line. However, this curve does not correlate with known sources or mechanisms, and none of the fits suggest a meaningful relation.

No significant linear relation between $PM_{10}$ and surface area was found at the Bountiful site, as idicated by a p-value of 0.094. The spline fit did fit some non-linearity, but the nature of the relation does not suggest a meaningful correlation. A small positive linear relation ($p = 0.048$) for $PM_{2.5}$ shows an average increase of 0.0016 μg/m³ of $PM_{2.5}$ for each additional square mile of surface area; however, again the nature of the relation and fit by the model suggests that it reflects only random fluctuations. No linear relation is seen in the coarse fraction ($p = 0.74$), and the non-linear spline fit also shows no relation ($p = 0.098$). At Hawthorne, a positive linear result for $PM_{10}$ ($p < 0.001$) with a small positive slope of 0.00025 μg/m³ means that every square mile of surface area gained is correlated with an increase of 0.00025 μg/m³ of $PM_{10}$. $PM_{2.5}$ at Hawthorne shows a stronger (but still small) positive relation, with a slope of 0.0026 ($p < 0.001$). The coarse fraction also shows a small positive slope of 0.005 μg/m³/miles² (p < 0.001). The Herriman site shows a slight negative trend of $PM_{10}$, with a slope of $-0.0012$ μg/$m^3$/$miles^2$ ($p < 0.001$), and $PM_{2.5}$ has a slope of $-0.0128$ μg/$m^3$/$miles^2$ ($p < 0.0001$); change in the coarse fraction does not reach statistical significance ($p = 0.156$). At the Magna site, the slope for $PM_{10}$ is 0.0004 μg/$m^3$/$miles^2$ ($p < 0.001$), but data for $PM_{2.5}$ do not have a significant slope ($p = 0.349$), nor do the data for the coarse fraction ($p = 0.117$). The Weber site $PM_{10}$ data show a small positive relation with surface area, slope = 0.00062 μg/$m^3$/$miles^2$ ($p < 0.001$). $PM_{2.5}$ also has a significant positive slope of 0.0021 μg/$m^3$/$miles^2$ ($p < 0.001$), and the coarse fraction has a slope of 0.006 μg/$m^3$/$miles^2$ ($p < 0.001$).

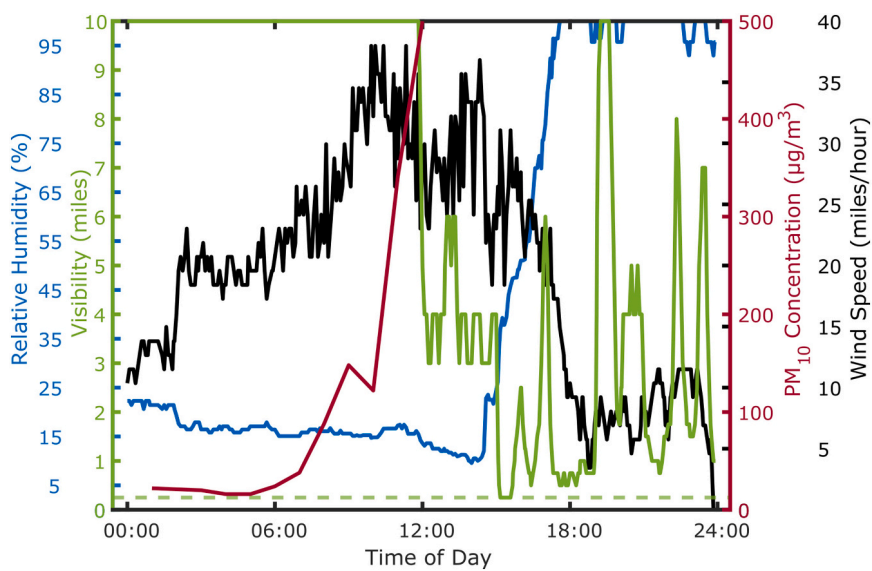

**Fig. 4.** An example of dust event detection algorithm output as a time series for the first captured dust event in Bountiful in October 2013. The dashed green line indicates a visibility of 0.25 miles.

Overall, we do not see any meaningful relation between the GSL surface area and $PM_{10}$, $PM_{2.5}$, or coarse fraction concentrations. Some small relations are statistically significant, but some are positive and some are negative and the curved relations show areas of both positive and negative relation. This is because spline fits pick up effects of unmeasured variables. Although we might be able to identify what these variables are and which would smooth the spline fit, this is unlikely to change the outcome for the variable of interest. Even with the likely overfitted spline models, the differences in predicted PM measures are small relative to the overall variation.

*3.2. Quantifying the number of dust events and potential dust sources*

*3.2.1. Meteorological analysis of filters*

The meteorological plots, like those in Fig. 4, were used to identify the time of the day during which visibility dropped. This time stamp was then used to identify which of a series of wind rose plots would best inform the direction from which the wind originated and where the potential dust sources were. Table 4 shows the representations of wind directions for each period (local time; UTC-6 or UTC-7) for which wind rose plots were made: night (00:00–08:00), morning (08:00–16:00), and afternoon (16:00–00:00). An example of these wind roses can be found in Fig. 5; the remainder can be found in supplemental materials.

These time series captures wind shifts from atmospheric-boundary passages on a continuum of spatial scales: synoptic (cold and warm fronts) (Schultz et al., 2002), mesoscale (lake breeze) (Blaylock et al., 2017), and smaller (canyon up- and down-slope flow) (Chrust et al., 2013).

The above analysis revealed that during days with "low" $PM_{10}$ concentrations, defined as filters that have a daily average concentration of <16 μg/$m^3$, the wind predominantly emanated from the north and west; however, winds were measured from almost every direction. Conversely, all winds linked to "high" concentration filters, defined as filters that have a concentration of >140 μg/$m^3$, exhibited a strong southerly component, with most of the wind coming from the SSE during times of day when $PM_{10}$ spiked. This indicates that elevated dust concentrations can occur when winds originate from the south. This finding is consistent with numerous referenced papers (Lang et al., 2023; Carling et al., 2020; Nicoll et al., 2020; Hahnenberger and Perry, 2015; Putman et al., 2022; Goodman et al., 2019; Bouton et al., 2020; Hahnenberger and Nicoll, 2012; Hahnenberger and Nicoll, 2014; Steenburgh et al., 2012; Munroe et al., 2023), which assert that significant dust contributions arise from the southern playas (Sevier Dry Lake, Tule Dry Lake, Milford Valley, Escalante Desert) and the western deserts (Great Salt Lake Desert, Dugway Proving Grounds, West Desert of Utah, Great Basin, and Mojave Deserts of Nevada), with possible minor contributions from the Great Salt Lake playa.

*3.2.2. Historical dust event analysis*

Meteorological data, including wind speed, RH, and visibility, were used to quantify the number of dust events at each of the five representative urban sites that surround the GSL. An example of the algorithm used to analyze the visibility data is in Fig. 4. Raw observation data were not consistently below the threshold per sample. This temporal variability motivated the form of smoothing and clustering discussed above. Extensive preliminary testing of parameter and methodology choice provided confidence in this clustering method, which aligns with the NWS definition (US National Weather Service, 2024) of a dust event. Qualifiers for a dust event included a wind speed of over 2 m/s (to rule out inversion events), a relative humidity (RH) below 95 % (to rule out any precipitation) and a visibility of 0.5 miles. Visibility of 0.5 miles was chosen to estimate the number of dust events found with this method. The meteorological data quantified a total number of dust events at each site as well as the average number of dust events per year at each site. These results are in Table 5.

Qualifying a dust event purely through meteorological data proved difficult. To enhance this analysis, hourly averaged $PM_{10}$ values were used to find dust events. Events that were within 3 h of each other were counted as single events, as per the NWS definition, and the number of hours that the $PM_{10}$ concentration was above 150 μg/$m^3$ was added up for each event. This second analysis was more robust to changes in meteorological analysis, giving an average of 17 dust events a year across the Wasatch Front with a total average duration of 98 min. These were the numbers used in the Health Risk Assessment. Results from this analysis for each of the three sites that had hourly averaged data are in Table 6.

**Table 4**
Representative wind directions for each period for each filter studied. Wind classified as "Low" is defined as 0–6 m/s, "Mod" is defined as 6 to 12 m/s, and high is defined as >12 m/s.

| Site | Date | Notes | Night | Morning | Afternoon |
|---|---|---|---|---|---|
| Herriman (H3) | 04/23/2022 | Baseline | Low NW | Low NW & SW | Low E & Mod NW |
| Herriman (H3) | 05/31/2022 | Baseline | Low E | Low SW | Low W to SW |
| Herriman (H3) | 06/17/2022 | 3rd from max | Mod SE | Low SE | Mod SE |
| Herriman (H3) | 06/20/2022 | Baseline | Mod SW | Low NW | Low SE |
| Utah Tech Center (EQ) | 06/17/2022 | 2nd from max | Mod SE | Low SE | Low SE |
| Utah Tech Center (EQ) | 06/18/2022 | 3rd from max | Mod S | Mod SE | Mod S |
| Utah Tech Center (EQ) | 06/20/2022 | Baseline | Mod SSW | Mod NNW | Low SSE |
| Utah Tech Center (EQ) | 09/07/2020 | Exceedance | Low N | Low SSE | Low SSE |
| Utah Tech Center (EQ) | 09/09/2020 | Baseline | High NE | Low SSE | Mod NNE |
| Herriman (H3) | 12/20/2017 | Exceedance | Mod SE | Mod SSE | Mod SSE |
| Magna (MG) | 12/20/2017 | Exceedance | Low NW | Low N | High S |
| Magna (MG) | 12/21/2017 | Baseline | High NNW | Mod NNW | Mod NNW |
| Herriman (He) | 04/14/2015 | Exceedance | Mod ESE | Low S | Mod SSE |
| Herriman (He) | 04/15/2015 | Baseline | Mod N & High S | Mod NW | Low WNW |
| Ogden (O2) | 04/14/2015 | Exceedance | Low S | Mod SSE | High S |
| Ogden (O2) | 04/15/2015 | Baseline | High NNW & SSW | Low N | Mod NW |

### 3.3. Dust compositional analysis and health risk assessment

Compositional $PM_{10}$ analysis of the dust collected on the filters reveals no significant increase in metal concentrations. In certain instances, the measured metal concentrations fell below the limit of detection of the ICP-MS method used for the analysis. Although lead levels were slightly below the quality assurance standards set by the EPA's methods, all other quality assurance checks and interference checks passed successfully.

The greatest amounts of Ba, Cd, Co, Cr, Mn, Ni, Pb, and V were detected on a filter collected at the Utah Tech Center (EPA AIRS code: 490353015) on June 17, 2022 (filter number: 0634559). The highest concentrations of Cu and Zn were observed on a filter from Herriman (H3) (EPA AIRS code: 490353013) on June 20, 2022 (filter number: 0634539). The highest levels of As were on a filter from Herriman (H3) (EPA AIRS code: 490353013) on May 31, 2022 (filter number: 0634365). Lastly, the highest levels of Hg were on a filter from the Utah Tech Center (EPA AIRS code: 490353015) on September 9, 2020. Notably, the filter from June 17, 2022, recorded a high concentration of $PM_{10}$ (146.7 μg/m$^3$), whereas those from June 20, 2022, May 31, 2022, and September 9, 2020, showed lower concentrations (15.2, 6.2, and 10.6 μg/m$^3$, respectively). Although metals were found in filters from Magna (EPA AIRS code: 49035101) and Ogden (EPA AIRS code: 490570002), these filters did not have the highest levels of any metals measured. Concentrations of metals in this work were compared to measured concentrations of metals at the UDAQ Bountiful site (Fig. 6).

The calculation of hazard quotients involved taking the highest concentrations of each metal to report a worst-case measured hazard quotient. A hazard quotient exceeding one suggests potential health risk and raises concerns. Conversely, hazard quotients below one indicate that the measured metal concentrations in the dust are of lesser concern. This analysis employed the maximum concentrations of metals in historical data to determine the most concerning hazard quotients. Table 7 presents the concentration of each metal, expressed in μg/m$^3$, along with their respective hazard quotients. Cancer Risk values were also calculated for the elements that had IUR values available, given in Table 7. Elements with a cancer risk value $>1 \times 10^{-6}$ indicate the need for an in-depth toxicological effects analysis (Agency for Toxic Substances and Disease Registry (ASTSDR), 2022). Similar calculations were made for the highest metal concentrations found on the filters analyzed in this paper and are presented in Table S1.

As indicated in Table 7, all hazard quotients are consistently below one. This implies minimal reason for concern about the concentration of any measured metals in the dust collected by UDAQ on their filters. Likewise, the cancer risk values are all $<1 \times 10^{-6}$. In other words, metals present in the dust reaching urban measurement sites are below established levels of concern.

EPA Air Regional Screening Levels (RSL) were compared to the highest concentrations of each metal reported, see Table 7. The RSLs are used to designate when an element becomes a chemical of potential concern (COPC) and merits further health risk assessment at a superfund site. Based on the highest concentration comparison As, Cd, Co, Mn, and Ni would be designated as COPC and require further health risk assessments. However, this assumes these concentrations are chronic exposures. These concentrations occur only infrequently during dust events, suggesting that further health risk assessments would lead to the conclusions above that the hazard quotient and cancer risk for all the metals are below levels of concern.

### 4. Conclusions

Recent concerns about the toxicity of dust from the exposed playa of the GSL have prompted a reevaluation of past research. In this study, we examined the dust that reached three urban monitoring sites surrounding the GSL, managed by the Utah Division of Air Quality. After examining historical data, we found no correlation between decreases in the surface area of the Great Salt Lake and increases in dust events in urban areas. We also found a general decrease in urban dust despite a decrease in the surface area of the GSL. This is consistent with some of the UDAQ data (United States Environmental Protection Agency, 2024a; United States Environmental Protection Agency, 2023a).

Wind rose plots align with previous research (Lang et al., 2023; Carling et al., 2020; Nicoll et al., 2020; Hahnenberger and Perry, 2015; Putman et al., 2022; Goodman et al., 2019; Bouton et al., 2020; Hahnenberger and Nicoll, 2012; Hahnenberger and Nicoll, 2014; Steenburgh et al., 2012; Munroe et al., 2023), indicating that elevated dust levels in urban areas coincide with winds originating from the south or west, passing over playas and deserts such as the Milford Flats, Sevier Dry Lake, Tule Dry Lake, Great Salt Lake Desert, Dugway Proving Grounds, and the West Desert of Utah.

A compositional analysis of urban dust revealed no concentrations of studied metals that raise health concerns. Air regional screening levels,

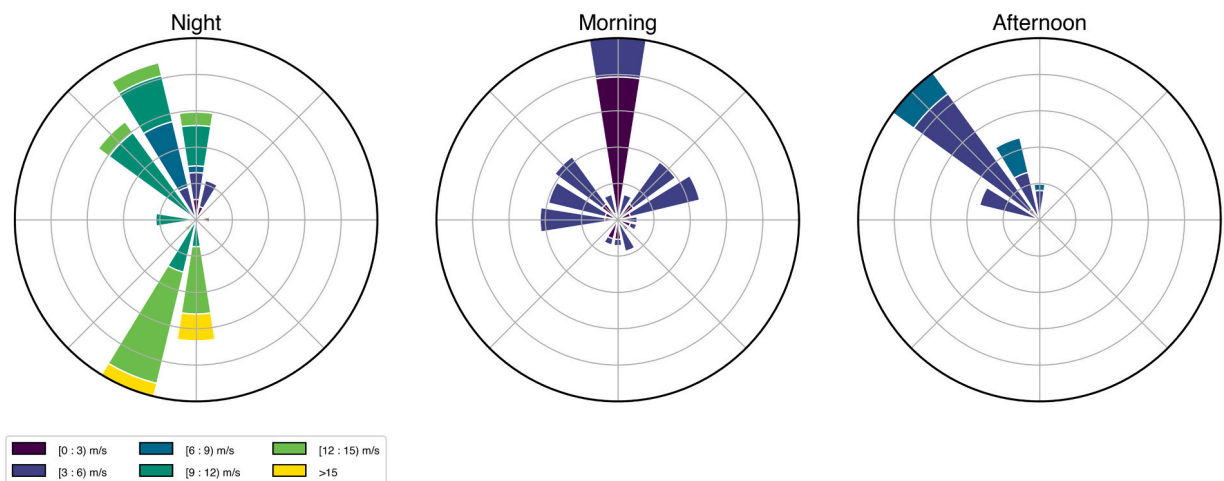

**Fig. 5.** Example of wind roses made for filter 2917235 from Ogden collected on 04/15/2015. The radial length indicates the percentage of that eight-hour period, and the color shows the wind-speed bin (meters per second).

**Table 5**
Sites used for meteorological data for each PM monitoring site and prediction of dust events.

| UDAQ site | Nearest Met STID | Distance to UDAQ site (km) | Years operational | Total number of dust events | Number of dust event per year | Average length of dust event (min) |
|---|---|---|---|---|---|---|
| BV | UTLGP | 2.8 | 5.5 | 12 | 2.2 | 112 |
| HW | UT20 | 4.4 | 8.5 | 24 | 2.5 | 59 |
| H3 | UTJUN | 3.6 | 9.5 | 27 | 2.7 | 77 |
| MG | KSLC | 13.0 | 10 | 3 | 0.4 | 33 |
| O2 | KOGD | 3.7 | 10 | 9 | 0.9 | 78 |

UTLGP is Legacy Parkway, Utah DoT (40.90772, −111.91657) UT20 is I-15 at 500 South WB (Gateway) (40.75829, −111.9121) KOGD – Ogden Hinckley Airport (41.19406–112.01681) UTJUN – SR-85 at Juniper (40.48607–11,199,415) KSLC – SLC International Airport (40.77069–11,196,503).

**Table 6**
Measured dust events and duration of exposure using hourly averaged $PM_{10}$ data.

| UDAQ site | Years operational | Total number of dust events | Number of dust event per year | Average length of dust event (min) |
|---|---|---|---|---|
| HW | 2022–2023 | 29 | 14.5 | 130 |
| H3 | 2012–2015 | 12 | 12 | 80 |
| O2 | 2023 | 73 | 24.3 | 84 |

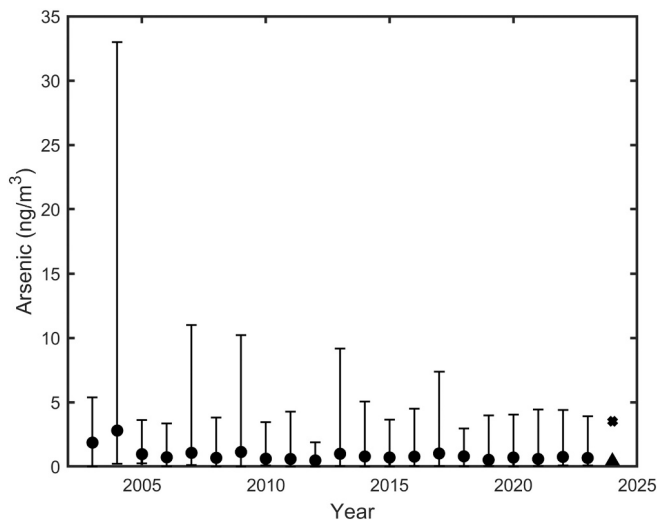

**Fig. 6.** Metal speciation plots of available metal concentrations found in historical measurements, reported as an annual average from 2003 to 2022, and in exceedance filter analysis, reported as year 2024. The circles represent annual averages provided by UDAQ with the minimum and maximum concentrations shown. Note that UDAQ reports metals every six days, so true maxima and minima may not be reported. All annual averages are also reported for the Bountiful site. The square indicates the maximum metal concentration measured (Utah Tech Center (EQ) EPA site number: 490353015 filter number: 0634559 date: June 17, 2022, a high concentration filter) and the triangle the minimum metal concentration measured (Magna (MG) EPA site number: 490351001 filter number: 7672720 date: 12/21/2017, a baseline filter) in this paper's filter analysis. These measurements come from multiple sample sites as listed.

hazard quotients, and cancer risks were evaluated for the highest recorded concentration for each of the studied metals to form this conclusion. Metal concentrations in the exceedance filters were also compared to historical metal concentrations measured at the Bountiful UDAQ site. This is mostly consistent with previous work by Attah et al. (Attah et al., 2024) who tested for 47 metals and found that As and Li were above the EPA's residential soil RSLs. This discrepancy for As may be from one or a combination of three possibilities. The first is that not all the soil from the GSL playa is being lofted into the air for long enough to make it to urban areas. Attah et al. forced their soil samples to aspirate to collect $PM_{10}$, which may not be representative of the ability of wind alone to resuspend dust from the GSL playa. Second, the dust generated from the GSL playa undergoes dilution by other dust and air volume before it makes its way to urban areas; therefore, the As levels are no longer in exceedance of health thresholds when they reach urban areas. Finally, Attah et al. used residential soil RSLs, not air RSLs as their qualifier in determining their RSLs for As (Attah et al., 2024).

Some limitation in data exists for speciated metal concentrations determined by UDAQ from the Bountiful site. Specifically, the state is required to do a filter analysis only once per six days, which could potentially miss dust storms. Also, many of the Hg concentrations in the

Table 7

The highest elemental concentrations recorded by UDAQ at BV or found in filters analyzed in this work and corresponding hazard quotients and cancer risk with corresponding air regional screening level (RSL). Note that no error limit is reported by UDAQ in their data.

| Element | Highest concentration (µg/m$^3$) | Hazard quotient | EPA RfC (µg/m$^3$) | Cancer Risk | EPA IUR (µg/m$^3$)$^{-1}$ | EPA RSL (µg/m$^3$) |
|---|---|---|---|---|---|---|
| As[a] | $3.3 \times 10^{-2}$ | $7.0 \times 10^{-3}$ | $1.5 \times 10^{-2}$ | $4.5 \times 10^{-7}$ | $4.3 \times 10^{-3}$ | $6.5 \times 10^{-4}$ |
| Ba[b] | $1.0 \times 10^{-1} \pm 5.6 \times 10^{-3}$ | $6.6 \times 10^{-4} \pm 4 \times 10^{-5}$ | $5.0 \times 10^{-1}$ | – | Lack of studies | $5.2 \times 10^{-1}$ |
| Cd[a] | $1.5 \times 10^{-2}$ | $4.8 \times 10^{-3}$ | $1.0 \times 10^{-2}$ | $8.7 \times 10^{-8}$ | $1.8 \times 10^{-3}$ | $1.6 \times 10^{-3}$ |
| Co[b] | $2.0 \times 10^{-3} \pm 1 \times 10^{-4}$ | $3.2 \times 10^{-4} \pm 2 \times 10^{-5}$ | $2.0 \times 10^{-2}$ | $5.7 \times 10^{-8} \pm 3 \times 10^{-9}$ | $9 \times 10^{-3}$ | $3.1 \times 10^{-4}$ |
| Cr[a] | $1.4 \times 10^{-2}$ | $4.5 \times 10^{-3}$ | Not available | – | Not available | Not available |
| Cu[b] | $3.8 \times 10^{-2} \pm 2 \times 10^{-3}$ | $1.2 \times 10^{-6} \pm 7 \times 10^{-8}$ | Not available | – | Not available | Not available |
| Hg[b] | $9.2 \times 10^{-4} \pm 5 \times 10^{-5}$ | $9.7 \times 10^{-6} \pm 5 \times 10^{-7}$ | $3.0 \times 10^{-1}$ | – | Inadequate data | $3.1 \times 10^{-1}$ |
| Mn[b] | $1.5 \times 10^{-1} \pm 8 \times 10^{-3}$ | $9.4 \times 10^{-3} \pm 5 \times 10^{-4}$ | $5.0 \times 10^{-2}$ | – | No human data | $5.2 \times 10^{-2}$ |
| Ni[a] | $3.0 \times 10^{-2}$ | $6.7 \times 10^{-3}$ | $1.4 \times 10^{-2}$ | $2.3 \times 10^{-8}$ | $2.4 \times 10^{-4}$ | $1.2 \times 10^{-2}$ |
| Pb[a] | $3.1 \times 10^{-2}$ | $6.6 \times 10^{-3}$ | Not available | – | Inadequate human data | $1.5 \times 10^{-1}$ |
| V[b] | $1.9 \times 10^{-2} \pm 1 \times 10^{-3}$ | $5.9 \times 10^{-4} \pm 3 \times 10^{-5}$ | $1.0 \times 10^{-1}$ | – | Not available | $1.0 \times 10^{-1}$ |
| Zn[b] | $1.2 \times 10^{-1} \pm 6 \times 10^{-3}$ | $7.6 \times 10^{-8} \pm 4 \times 10^{-9}$ | Not available | – | Inadequate data | Not available |

[a] Data from UDAQ (BV filters).
[b] Data from filters analyzed in this paper.

exceedance filters fell below the method detection limit. However, this implies that all concentrations of Hg in dust samples were below health risk levels.

Future work may include studying whether dust events at each site are increasing or decreasing in future. This leverage hourly averaged data now provided by UDAQ using the same definitions found in this work. Another fruitful avenue of study may include the influence of using water from the Utah and Bear Lakes to fill up the GSL. An analysis of dust, or potential dust, coming off these lakes might be studied if greater exposed playa results from extraction of water from these lakes into the GSL.

**CRediT authorship contribution statement**

**Callum E. Flowerday:** Writing – review & editing, Writing – original draft, Methodology, Formal analysis, Data curation, Conceptualization. **Rebekah S. Stanley:** Writing – review & editing, Methodology, Formal analysis, Data curation. **John R. Lawson:** Writing – review & editing, Methodology, Funding acquisition, Formal analysis, Data curation. **Gregory L. Snow:** Writing – review & editing, Methodology, Formal analysis, Data curation. **Kaitlyn Brewster:** Methodology, Formal analysis, Data curation. **Steven R. Goates:** Writing – review & editing, Methodology, Formal analysis. **Walter F. Paxton:** Writing – review & editing, Funding acquisition. **Jaron C. Hansen:** Writing – review & editing, Supervision, Methodology, Funding acquisition, Formal analysis, Conceptualization.

**Funding**

This study was funded by the National Science Foundation, grant #2114655. The funder played no role in the study design, data collection, analysis and interpretation of data, or the writing of this manuscript. JRL was supported by *Uintah County Special Service District 1* and the Utah Legislature.

**Declaration of competing interest**

All authors declare no financial or non-financial competing interests.

**Acknowledgements**

Young Living – Use of their ICP-MS for analysis. Utah Division of Air Quality – providing sample filters for analysis.

**Appendix A. Supplementary data**

Supplementary data to this article can be found online at https://doi.org/10.1016/j.scitotenv.2024.178202.

**Data availability**

The datasets generated and/or analyzed during the current study are available in the BYU's Scholar's Archive repository, https://scholarsarchive.byu.edu/data/62, and https://github.com/Bingham-Research-Center/gsl-dust.